# Evaluation of microstructure and mechanical property variations in Al$_x$CoCrFeNi high entropy alloys produced by a high-throughput laser deposition method

Mu Li[1], Jaume Gazquez[2], Albina Borisevich[3], Rohan Mishra[1,4], and Katharine M. Flores[1,4]

1. Mechanical Engineering and Materials Science, Washington University in St. Louis, St. Louis, MO 63130, USA
2. Institut de Ciència de Materials de Barcelona, Barcelona 08193, Spain
3. Materials Science and Technology Division, Oak Ridge National Laboratory, Oak Ridge, TN 37831, USA
4. Institute of Materials Science and Engineering, Washington University in St. Louis, St. Louis, MO 63130, USA

## Abstract

Twenty-one distinct Al$_x$CoCrFeNi alloys were rapidly prepared by laser alloying an equiatomic CoCrFeNi substrate with Al powder to create an alloy library ranging x = 0.15 – 1.32. Variations in crystal structure, microstructure and mechanical properties were investigated using X-ray diffraction, scanning electron microscopy, scanning transmission electron microscopy and nanoindentation. With increasing Al content, the crystal structure transitioned from a disordered FCC to a mixture of disordered BCC and ordered B2 structures. While the onset of BCC/B2 formation was consistent with previously reported cast alloys, the FCC structure was observed at larger Al contents in the laser processed materials, resulting in a wider two phase regime. The FCC phase was primarily confined to the BCC/B2 grain boundaries at these high Al contents. The nanoindentation modulus and hardness of the FCC phase increased with Al content, while the properties of the BCC/B2 structure were insensitive to composition. The structure and mechanical properties of the laser-processed alloys were surprisingly consistent with reported results for cast alloys, demonstrating the feasibility of applying this high-throughput methodology to multicomponent alloy design.

## Introduction

High entropy alloys (HEAs), also referred to as multi-principle element alloys, are of growing interest as structural materials. HEAs are typically defined as random solid solutions consisting of 5 or more metallic elements without a dominant solvent species [1]. This compositional complexity is often combined with a surprising simplicity in the crystal structure, including disordered face centered cubic (FCC), body centered cubic (BCC), and hexagonal close packed (HCP) structures [2]. Although the formation of single-phase HEAs is thought to result from the competition between the enthalpy of formation of intermetallic compounds and entropy of mixing of the large number of components in the alloy, identifying alloys which form these relatively simple structures within the vast multicomponent composition space presents a significant challenge. To address this, in the present work we apply a high-throughput laser deposition method [3–5] to efficiently create compositional libraries in the Al$_x$CoCrFeNi system. Compositional trends in microstructure and mechanical behavior are then evaluated.

Among the most studied systems, Al-transition metal HEAs such as Al$_x$CoCrFeNi and Al$_x$CoCrCuFeNi, have been found to have superior high-temperature strength as well as wear and oxidation resistance [6–8]. The unexpected properties beyond the capabilities of the constituent elements can be attributed to the presence of multiple principal elements, which are believed to induce in a range of complex effects, including lattice distortion and sluggish diffusion [9,10]. Al$_x$CoCrFeNi has been shown to transition from FCC to BCC with increasing Al content [11–14], making it of particular interest for a study of microstructural effects on properties.

Conventionally, to investigate phase formation and microstructure of complex metallic alloys, researchers adopt a serial approach to prepare separate samples of each composition of interest, typically via casting, melt spinning or powder metallurgy techniques. This requires significant investment of time and resources. Laser-aided direct metal deposition provides a way to produce extensive alloy libraries with much greater efficiency [3–5]. The energy provided by the laser creates a melt pool on the surface of a substrate, which can be a pure substance or prealloyed. A metallic powder stream is then delivered to the melt pool to alloy the material. By varying the powder composition and feed rate, a wide range of compositions can be created on a single substrate. In this work, we apply this methodology to Al$_x$CoCrFeNi. Using the laser-processed alloy libraries, the evolution of the microstructure is characterized by X-ray diffraction and electron microscopy, and mechanical properties are measured via nanoindentation. The results are generally in good agreement with those obtained from a serial casting approaches reported in the literature, while requiring significantly less time to produce. This demonstrates the feasibility of applying the high-throughput methodology to the design of complex crystalline alloys.

**Materials and Methods**

An equiatomic CoCrFeNi substrate was prepared using arc melting and casting. Raw materials with purity greater than 99.9% were melted together in an argon atmosphere. The resulting ingot was flipped and remelted at least 5 times on the water-cooled copper hearth to improve homogeneity prior to casting into a copper mold to produce a 25 mm × 25 mm × 4.5 mm plate.

Compositional libraries were prepared using an Optomec MR-7 LENS™ system. Twenty-five 2 mm × 2 mm patches on the substrate surface were alloyed with varying amounts of aluminum powder. The library had a total size of 14 mm × 14 mm. In the Laser Engineered Net Shaping (LENS) process, the laser creates a melt pool on the substrate surface, into which a stream of the alloying powder is directed. The laser and powder stream raster over the surface to create a layer; subsequent layers may be added to create a three-dimensional deposit if desired. In the present experiments, the laser power and travel speed during deposition were held constant at 150 W and 12.7 mm/s (30 in/min), respectively, while the Al powder feed rate increased from 0.8 to 3.3 rpm in increments of 0.1 rpm for each single-layer patch. Each patch consisted of 5 laser tracks with approximately 25% overlap. The patches were remelted twice perpendicular to the deposition direction at a laser power of 200 W to improve mixing and compositional homogeneity of the alloyed region. The entire deposition process to create the compositional library of 25 patches was completed in less than 15 minutes. The surface of the library was then cleaned, ground, and polished for microstructural characterization.

The crystal structure of each patch was characterized via X-ray diffraction (XRD) with a Rigaku D-MAX/A diffractometer with Cu-K$_\alpha$ radiation operating at 35 kV / 35 mA. Because the spot size of the X-ray beam on the sample surface was larger than the 2 mm x 2 mm square patches, a plexiglass mask with a tapered circular aperture (minimum diameter 2 mm) was employed to block the signal from the surrounding material, enabling each patch to be isolated for characterization. The diffraction signal from the mask was normalized and subtracted from the XRD patterns.

The microstructures of the patches were characterized using a JEOL JSM-7001FLV field emission scanning electron microscope (SEM) equipped with an Oxford HKL electron backscattered diffraction (EBSD) detector. To most efficiently characterize the alloy library, all of the SEM and related observations were performed in the plan view, i.e. perpendicular to the substrate surface, rather than in cross section. The compositions of the patches were measured using an Oxford INCA Energy 350 Si (Li) energy dispersive x-ray spectroscopy system (EDS) on the SEM. For each sample, random points in 5 different grains were evaluated.

In order to evaluate the effect of composition on mechanical properties, the hardness and elastic modulus of each patch were measured using a Hysitron TI 950 Triboindenter equipped with a diamond Berkovich indenter tip. Nanoindentation was performed on the polished plan view surface at a loading and unloading rate of 600 μN/s up to a maximum load of 3000 μN. Four indents were performed on each patch or, when possible, within each phase of interest, with a spacing of at least 20 μm between indents. Load-displacement data was continuously captured and used to determine the modulus, E, and hardness, H, using the method described by Oliver and Pharr [15] and assuming a constant value of Poisson's Ratio of 0.25 for Al$_x$CoCrFeNi.

Finally, detailed scanning transmission electron microscopy (STEM) was performed for a selected composition using a Nion UltraSTEM 200 microscope, operating at 200 kV and equipped with a fifth-order Nion aberration corrector and a cold-field emission gun. Due to the destructive nature of STEM specimen preparation, an Al$_{0.62}$CoCrFeNi specimen was prepared separately from the main compositional library, using the same LENS parameters outlined above. The patch used for the STEM analysis was 2 mm × 5 mm. The STEM specimen was prepared using conventional grinding, polishing and ion milling methods. High angle annular dark field (HAADF) Z-contrast imaging was performed using a probe convergence angle of 30 mrad and an annular dark-field detector with an inner angle larger than 86 mrad. Electron energy loss spectroscopy (EELS) maps were collected with a Gatan Enfinium EEL spectrometer in the STEM using a collection semi-angle of 33 mrad with a dwell-time of 0.2 sec/pixel. To reduce random noise in the EELS maps, principle component analysis (PCA) was performed.

**Results**

The range of Al powder feed rates and processing conditions used in these experiments resulted in 21 distinct Al$_x$CoCrFeNi alloys with Al contents that vary from x = 0.15 to 1.32 (3.73 - 24.78 at.% Al). The Al fraction increases linearly with the powder feed rate, as shown in Figure 1, indicating that the feed rate can be calibrated for use as a reasonable indicator for the final composition. In the remaining discussion, the alloy compositions will be identified by the stoichiometric subscript value x.

As expected, the crystal structure of $Al_xCoCrFeNi$ depends on Al content, as described in Table 1. The X-ray diffraction patterns of 7 selected compositions are shown in Figure 2. Peaks corresponding to FCC, BCC, and ordered B2 crystal structures are identified. For low Al contents, $x \leq 0.37$, only the FCC structure is observed. The lattice parameter for FCC increases slightly from 3.59 Å (x = 0) to 3.61 Å (x ≥ 0.69). A disordered BCC structure appears at x = 0.41, with a superlattice peak corresponding to an ordered B2 phase appearing when x ≥ 0.52. The superlattice peak shifts with increasing Al content, from 2θ = 30.56° for x = 0.52 to 2θ = 31.30° for x = 1.16, corresponding to a decrease in the B2 lattice parameter from 2.92 to 2.85 Å with the Al addition. The lattice parameter of the disordered BCC structure remains constant at 2.88 Å. At the highest Al contents examined, x = 1.16 and 1.32, the FCC peaks are no longer observed, and only the BCC/B2 structure remains.

For all of the compositions investigated, laser processing resulted in high aspect ratio microstructural features elongated to the direction of the laser path, with feature lengths up to 100 μm and widths on the order of tens of microns. Figures 3 and 4 show the evolution of the laser-processed microstructures with increasing Al content in more detail. For aluminum contents up to x = 0.37, the microstructure of the single FCC phase is very uniform, without evidence of any compositional fluctuations or segregation (Figure 3a). A second phase becomes apparent at x = 0.41, as seen in Figure 4a, where a lighter boundary-phase creates a "cellular" structure surrounding the darker majority phase. EBSD results shown in Figure 4c indicate that this new phase is BCC, consistent with the XRD results. (Regions with the B2 structure could not be differentiated from the BCC.) The orientation map shows that the cellular structures are arranged in colonies of the same orientation, with the BCC phase occurring both within and at the edges of the colonies. As the Al content increases, the areal fraction of the BCC/B2 structure increases linearly from 11.1% at x = 0.41, to 26.1% at x = 0.54. An EDS line scan across the cellular features in an alloy with x = 0.41, shown in Figure 5, indicates that the BCC/B2 boundary phase is enriched in Al.

The BCC/B2 structure becomes the dominant phase when x = 0.69. At this composition, secondary electron imaging, shown in Figure 4d, reveals a mottled grain structure consisting of dendrites surrounded by cell-like walls. These variations are also apparent at higher Al contents using electron backscatter imaging, as shown in Figure 3c and 3d. Figure 3e shows the interior of the dendrites at x = 1.16 in greater detail, revealing spherical particles embedded in the matrix. Figure 3f, obtained from the interdendritic region for the same composition, shows a weave-like structure typical of spinodal decomposition, consistent with prior observations [14]. Figure 4e and 4f shows the EBSD results for x = 0.69. The grains are uniformly BCC/B2, with a small amount (approximately 3.6% area) of the FCC phase present along the grain boundaries.

To further investigate the BCC/B2 structure, we performed atomic-scale characterization using STEM imaging and EELS. The composition library was supplemented with an additional laser processed specimen (x = 0.62) sacrificed for STEM analysis. Figure 6a shows a HAADF Z-contrast STEM image and corresponding diffraction patterns obtained using

Fourier transformation of the selected region of the image. The BCC and B2 lattices are coherent at this Al content. EELS maps reveal strong compositional segregation between the two microconstituents, as shown in Figure 6b. The B2 microconstituent is enriched in Al, Co and Ni, while the BCC is enriched in Fe and Cr. We did not observe any FCC phase in the small volume of material characterized in STEM, which confirms the small fraction (~3.6 %) of FCC phase as obtained from XRD analysis and their strong spatial inhomogeneity with a tendency to segregate to grain boundaries based on the EBSD results.

Several prior studies of $Al_x$CoCrFeNi have examined the bulk mechanical properties [11,13,14], as well as the nanoindentation hardness and modulus for a limited number of compositions [12,16–18]. In the present work we use the localized nature of nanoindentation to measure the elastic modulus and hardness of the FCC and BCC/B2 phases individually with increasing Al content over the entire compositional library. The results are presented in Figure 7. Note that, although both FCC and BCC/B2 structures are observed over a wide composition range (denoted by the dotted boundaries in the plot), the small dimensions of the FCC regions for x > 0.54 precluded isolating those regions for nanoindentation. However, independent measurements of both phases were made for x = 0.48, 0.52, and 0.54. Both the hardness and modulus are observed to increase linearly with increasing Al content in the FCC structure, while both remain reasonably constant with increasing Al content in the BCC/B2 structure. Data from the FCC-only (low Al content) and BCC/B2-only (high Al content) composition regions are consistent with the trends observed within the FCC + BCC/B2 region.

**Discussion**

Prior studies of $Al_x$CoCrFeNi system using serially cast materials have noted that the system transitions from FCC to a BCC/B2 structure with increasing Al content, with both phases forming over an intermediate composition range. Kao et al. [13] examined 10 compositions ranging x = 0 – 2. They observed a single FCC phase for x ≤ 0.34, two phase structures (FCC + BCC) for 0.57 ≤ x ≤ 0.71, and a single BCC phase for x ≥ 0.77. Similarly, Wang et al. [11] examined 13 compositions over the x = 0 – 2 range. They observed a single FCC phase for x ≤ 0.4, the addition of the BCC phase at x = 0.5, and the development of B2 at x = 0.7. Only the BCC/B2 structure was observed for 0.9 ≤ x ≤ 2.0. Ma et al. [14] examined 6 compositions, x = 0.41 – 1.33, and observed a "minor" amount of BCC phase in an FCC matrix at x = 0.41, the appearance of B2 at x = 0.57, and a BCC/B2 matrix with "minor" amounts of FCC at x = 0.74. Only the BCC/B2 structure was observed for x ≥ 0.92.

The 25 patches in the present laser-deposited compositional library include 21 distinct compositions over the range x = 0.15 – 1.32, providing the opportunity to examine the structural transitions in greater detail. Under the laser processing conditions used here, the BCC phase is first observed to precipitate from the FCC matrix at x = 0.41, consistent with the prior results, as Al is rejected from the FCC dendrites (Figure 5). Note that the BCC/B2 is found at FCC grain boundaries as well as within the grains, as shown in Figure 4. The B2 phase is first observed at x = 0.52, again consistent with both Ma et al. (who did not consider x values between 0.41 and 0.57) and Wang et al. (who did not consider x values between 0.5 and 0.7). The precipitation of the ordered B2 structure from the disordered BCC phase with increasing Al content is due to the larger negative bond enthalpies for Al-Ni (-22 kJ/mol) and Al-Co (-19 kJ/mol), in comparison with those for Al-Cr (-10 kJ/mol) and Al-Fe (-11 kJ/mol) [19]. This compositional segregation is apparent in the STEM-EELS data presented in Figure 6, where the B2 region is enriched in both Ni and Co and depleted in Cr and Fe compared with the BCC.

The FCC phase is still present in the laser-processed alloys at x = 1.06 (21 at.% Al), 2-5 at.% Al higher than the earlier studies using cast materials. The retained FCC phase is most likely the result of compositional fluctuations associated with the inherently high quench rate under laser processing conditions, which are not observed in the casting techniques used in the prior studies. While the FCC phase is still observed in the XRD data at these higher Al contents, it is only present in small amounts; recall that only 3.6% of the microstructure was identified as FCC at x = 0.69 (Figure 4b). This latter observation is consistent with Ma et al.'s observation of a "minor" amount of FCC at x = 0.74. Furthermore, the FCC phase is only found at isolated positions along the high-angle twist and tilt BCC/B2 grain boundaries, where the more open structure may enable the FCC lattice to accommodate additional Al.

Several authors have examined the ability of various simple parameters to predict phase stability in HEAs. Guo et al. [20] compared the valence electron concentration, VEC, the atomic size mismatch, $\delta$, and the electronegativity difference, $\Delta\chi$, for several HEAs (although notably, not $Al_x$CoCrFeNi), and concluded that VEC provided the best indication of the stability of the BCC and FCC phases. Based on empirical observations, they proposed that FCC would be the only phase present for VEC ≥ 8.0, and the BCC/B2 structure would be the only phase for VEC < 6.87. Intermediate VEC values would include a mixture of the two phases. These three parameters are calculated for the compositions in the present work (using the definitions and elemental values found in Guo et al. [20]) and compared with those for the compositions examined in the prior studies in Figure 8 [11,13,14]. While the two phase regime sits within Guo et al.'s predicted range, all 4 sets of data exhibit narrower FCC+BCC/B2 ranges than predicted. The two phase regime in the present work spans the widest composition range (as discussed above), most closely approaching Guo et al.'s prediction. This is quite surprising, given the highly non-equilibrium cooling of the laser processed material.

The laser-processed library provides the opportunity to study mechanical property variations with composition in detail via nanoindentation. Furthermore, the localized nature of nanoindentation measurements allows us to consider the FCC and BCC/B2 structures independently, including within the two phase regime. As shown in Figure 7a, the modulus for the BCC/B2 structure was insensitive to the Al content, with an average value of 187 GPa. This is consistent with the value reported for x = 1.5 [12], but is low compared with the value of 230 GPa for x = 1 reported elsewhere [16]. The modulus of the FCC structure exhibits an increasing trend, ranging 150-223 GPa in the present work. While the limited number of moduli values reported in the literature for low Al contents increase more slowly and are somewhat higher than those observed here [12,16], the reported range is again broadly consistent with the current measurements.

Some variation in the modulus may be expected due to differences in the orientation of the crystals being indented (which was not characterized here). However, one would expect orientational effects to result in increased variability within each composition, rather than the observed linear trend with Al content. The increasing trend in the modulus at low Al content may be attributed to Al's high electron density in the outermost shell and correspondingly high Fermi level. Al tends to transfer electrons to the transition metals, leading to a stronger atomic binding force for Al-TM bonds than TM-TM [21], as evidenced by bond-shortening and larger interatomic potentials [22,23], and thus increasing the bond stiffness. Although the lattice parameter of the FCC phase was observed to increase by 0.5%, from 3.59 Å (x = 0) to 3.61 Å (x ≥ 0.69), which should reduce the modulus, it increases far more slowly than the average of the atomic radii of the components (which increases by 2.2%), especially at low Al concentration. On balance, the increasing bond stiffness and more slowly increasing bond length results in an overall increase of the FCC elastic modulus with Al content.

Prior studies have also measured the variation in the hardness and compressive deformation behavior of cast $Al_x$CoCrFeNi with Al content [12–14,16]. The BCC/B2 structure is generally found to have higher hardness and strength than the FCC phase, due to the reduced number of dislocation slip systems in the BCC/B2 structure. While only a limited number of compositions have been characterized in the literature, prior nanoindentation measurements on cast alloys indicate that the hardness increases significantly with Al content. Tian et al. indicated that the hardness increases from approximately 3 GPa at x = 0.3 to almost 9 GPa at x = 1 [16], consistent with the data shown in Figure 7b. This agreement is somewhat surprising, since one would expect that the rapid quenching inherent to laser processing would result in significant thermal stresses, which in turn should cause a shift in the hardness values. Tian et al. also noted that $Al_x$CoCrFeNi exhibits significant strain rate sensitivity; the present measurements were performed at the same loading rate as that work. Other nanoindentation [12] and microhardness [13,14] measurements revealed somewhat lower hardness values for both phases, potentially due to the use of a lower loading rate.

While the present results are generally consistent with those of Tian et al., the compositional library permits a more detailed study of variations with composition. The nanoindentation hardness of the FCC structure increases slightly with the amount of Al. This can be attributed to solid solution strengthening, as the addition of larger Al atoms increases the lattice distortion energy. In contrast, the hardness of the BCC/B2 structure appears to be insensitive to the Al content, even though the precipitation of the ordered B2 structure would be expected to increase the strength relative to the disordered BCC phase. While the volume fraction of B2 may be expected to increase with Al content, the coherency strain due to lattice mismatch goes through a minimum, with the B2 phase switching from a state of compression to tension relative to the BCC at approximately x = 0.9; the low coherency strain at intermediate compositions reduces the effectiveness of precipitation strengthening. Different B2/BCC morphologies are also observed, even within a single composition as shown in Figure 3e and f, further complicating the analysis. This combination of several competing factors apparently results in the nearly constant hardness of the BCC/B2 structure with Al content. More work is required to compare the behavior of the different morphologies and, if possible, develop microstructural design strategies that will enhance the performance.

**Conclusions**

Microstructural and mechanical property variations in the $Al_x$CoCrFeNi system were successfully characterized using a high-throughput laser deposition-based fabrication technique. While the laser processed FCC phase was retained at higher than expected Al contents, resulting in a wider multiphase (FCC + BCC/B2) compositional range, the observed microstructures were quite consistent with those reported in the literature for serially cast materials. The moduli and hardness, obtained via nanoindentation over the entire composition range, were also in very good agreement with the limited number of reported measurements for cast materials obtained under similar loading conditions. This remarkable consistency between the cast and laser-processed microstructures and properties demonstrates that the high-throughput, laser deposition-based synthesis technique may be applied to the study of complex, multicomponent crystalline alloys, and that the structure and properties of the resulting compositional libraries are representative of materials prepared by conventional methods. By dramatically reducing the synthesis time to on the order of a minute per composition, this method has the potential to rapidly accelerate the rate at which the most promising compositions can be identified and refined, prior to bulk-scale preparation and testing.

**Acknowledgements**

The authors acknowledge financial support from Washington University in St. Louis and the Institute of Materials Science and Engineering for the use of the SEM and XRD and for staff assistance. STEM work performed at Oak Ridge National Laboratory was supported by the U.S. Department of Energy, Office of Science, Basic Energy Sciences Materials Science and Engineering Division (BES-MSED). JG acknowledges the Ramón y Cajal program for support (RyC-2012-11709).

**References**

[1] B. Cantor, Multicomponent and high entropy alloys, Entropy. 16 (2014) 4749–4768. doi:10.3390/e16094749.

[2] D.B. Miracle, O.N. Senkov, A critical review of high entropy alloys and related concepts, Acta Mater. 122 (2017) 448–511. doi:10.1016/j.actamat.2016.08.081.

[3] P. Tsai, K.M. Flores, A Laser Deposition Strategy for the Efficient Identification of Glass-Forming Alloys, Metall. Mater. Trans. A Phys. Metall. Mater. Sci. 46 (2015) 3876–3882. doi:10.1007/s11661-015-2900-x.

[4] P. Tsai, K.M. Flores, A combinatorial strategy for metallic glass design via laser deposition, Intermetallics. 55 (2014) 162–166. doi:10.1016/j.intermet.2014.07.017.

[5] P. Tsai, K.M. Flores, High-throughput discovery and characterization of multicomponent bulk metallic glass alloys, Acta Mater. 120 (2016) 426–434. doi:10.1016/j.actamat.2016.08.068.

[6] C.-J. Tong, Y.-L. Chen, J.-W. Yeh, S.-J. Lin, S.-K. Chen, T.-T. Shun, C.-H. Tsau, S.-Y. Chang, Mechanical Performance of the AlxCoCrCuFeNi High-Entropy Alloy System with Multiprincipal Elements, Metall. Mater. Trans. A. 36 (2005) 881–893. doi:10.1007/s11661-005-0283-0.

[7] J.M. Wu, S.J. Lin, J.W. Yeh, S.K. Chen, Y.S. Huang, H.C. Chen, Adhesive wear behavior of AlxCoCrCuFeNi high-entropy alloys as a function of aluminum content, Wear. 261 (2006) 513–519. doi:10.1016/j.wear.2005.12.008.

[8] P. Huang, J. Yeh, Multi-Principal Element Alloys with Improved Oxidation and Wear Resistance for Thermal Spray Coating, Adv. Eng. Mater. 6 (2004) 74–78.

doi:10.1002/adem.200300507.

[9] Y. Zhang, T.T. Zuo, Z. Tang, M.C. Gao, K.A. Dahmen, P.K. Liaw, Z.P. Lu, Microstructures and properties of high-entropy alloys, Prog. Mater. Sci. 61 (2014) 1–93. doi:10.1016/j.pmatsci.2013.10.001.

[10] J.W. Yeh, Recent progress in high-entropy alloys, Ann. Chim. Sci. Des Mater. 31 (2006) 633–648. doi:10.3166/acsm.31.633-648.

[11] W.-R. Wang, W.-L. Wang, S.-C. Wang, Y.-C. Tsai, C.-H. Lai, J.-W. Yeh, Effects of Al addition on the microstructure and mechanical property of AlxCoCrFeNi high-entropy alloys, Intermetallics. 26 (2012) 44–51. doi:10.1016/j.intermet.2012.03.005.

[12] T. Yang, S. Xia, S.S.S. Liu, C. Wang, S.S.S. Liu, Y. Zhang, J. Xue, S. Yan, Y. Wang, Effects of AL addition on microstructure and mechanical properties of AlxCoCrFeNi High-entropy alloy, Mater. Sci. Eng. A. 648 (2015) 15–22. doi:10.1016/j.msea.2015.09.034.

[13] Y.-F.F. Kao, T.-J.J. Chen, S.-K.K. Chen, J.-W.W. Yeh, Microstructure and mechanical property of as-cast, -homogenized, and -deformed AlxCoCrFeNi (0≤x≤2) high-entropy alloys, J. Alloys Compd. 488 (2009) 57–64. doi:10.1016/j.jallcom.2009.08.090.

[14] Y. Ma, B. Jiang, C. Li, Q. Wang, C. Dong, P. Liaw, F. Xu, L. Sun, The BCC/B2 Morphologies in AlxNiCoFeCr High-Entropy Alloys, Metals (Basel). 7 (2017) 57. doi:10.3390/met7020057.

[15] W.C. Oliver, G.M. Pharr, An improved technique for determining hardness and elastic modulus using load and displacement sensing indentation experiments, J. Mater. Res. 7 (1992) 1564–1583. doi:10.1557/JMR.1992.1564.

[16] L. Tian, Z.M. Jiao, G.Z. Yuan, S.G. Ma, Z.H. Wang, H.J. Yang, Y. Zhang, J.W. Qiao, Effect of Strain Rate on Deformation Behavior of AlCoCrFeNi High-Entropy Alloy by Nanoindentation, J. Mater. Eng. Perform. 25 (2016) 2255–2260. doi:10.1007/s11665-016-2082-8.

[17] Z.M. Jiao, Z.H. Wang, R.F. Wu, J.W. Qiao, Strain rate sensitivity of nanoindentation creep in an AlCoCrFeNi high-entropy alloy, Appl. Phys. A Mater. Sci. Process. 122 (2016) 1–5. doi:10.1007/s00339-016-0339-6.

[18] Z.M. Jiao, M.Y. Chu, H.J. Yang, Z.H. Wang, J.W. Qiao, Nanoindentation characterised plastic deformation of a $Al_{0.5}$CoCrFeNi high entropy alloy, Mater. Sci. Technol. 31 (2015) 1244–1249. doi:10.1179/1743284715Y.0000000048.

[19] M.C. Troparevsky, J.R. Morris, P.R.C. Kent, A.R. Lupini, G.M. Stocks, Criteria for predicting the formation of single-phase high-entropy alloys, Phys. Rev. X. 5 (2015). doi:10.1103/PhysRevX.5.011041.

[20] S. Guo, C. Ng, J. Lu, C.T. Liu, Effect of valence electron concentration on stability of fcc or bcc phase in high entropy alloys, J. Appl. Phys. 109 (2011). doi:10.1063/1.3587228.

[21] Z. Tang, M.C. Gao, H. Diao, T. Yang, J. Liu, T. Zuo, Y. Zhang, Z. Lu, Y. Cheng, Y. Zhang, K.A. Dahmen, P.K. Liaw, T. Egami, Aluminum alloying effects on lattice types, microstructures, and mechanical behavior of high-entropy alloys systems, JOM. 65 (2013) 1848–1858. doi:10.1007/s11837-013-0776-z.

[22] H.Y. Hsieh, B.H. Toby, T. Egami, Y. He, S.J. Poon, G.J. Shiflet, Atomic structure of amorphous Al90FexCe 10−x, J. Mater. Res. 5 (1990) 2807–2812. doi:10.1557/JMR.1990.2807.

[23] M. Widom, I. Al-Lehyani, J. Moriarty, First-principles interatomic potentials for transition-metal aluminides. III. Extension to ternary phase diagrams, Phys. Rev. B. 62 (2000) 3648–3657. doi:10.1103/PhysRevB.62.3648.

**Figures and Tables**

**Table 1**

Measured compositions and structures of 25 laser processed patches, and the nominal compositions of the corresponding $Al_xCoCrFeNi$ alloy.

| Measured Composition | | | | | Nominal Composition | Phase Structure |
|---|---|---|---|---|---|---|
| Al (at.%) | Co (at.%) | Cr (at.%) | Fe (at.%) | Ni (at.%) | | |
| 3.73 ±0.54 | 24.16 ±0.53 | 24.75 ±0.09 | 23.15 ±0.36 | 24.20 ±0.96 | $Al_{0.15}CoCrFeNi$ | FCC |
| 4.68 ±0.24 | 23.00 ±0.21 | 24.17 ±0.10 | 24.55 ±0.14 | 23.60 ±0.65 | $Al_{0.20}CoCrFeNi$ | FCC |
| 5.66 ±0.04 | 23.13 ±1.44 | 24.82 ±0.29 | 23.49 ±0.67 | 22.90 ±1.84 | $Al_{0.24}CoCrFeNi$ | FCC |
| 5.70 ±0.32 | 22.77 ±0.14 | 24.71 ±0.27 | 22.88 ±0.46 | 23.94 ±0.45 | $Al_{0.24}CoCrFeNi$ | FCC |
| 6.55 ±0.45 | 23.27 ±0.61 | 24.35 ±0.33 | 22.98 ±0.23 | 22.86 ±0.04 | $Al_{0.28}CoCrFeNi$ | FCC |
| 6.56 ±1.03 | 22.75 ±1.10 | 25.13 ±0.75 | 23.14 ±0.51 | 22.43 ±1.00 | $Al_{0.28}CoCrFeNi$ | FCC |
| 7.90 ±0.43 | 22.69 ±1.05 | 23.39 ±0.22 | 22.52 ±0.23 | 23.49 ±0.75 | $Al_{0.34}CoCrFeNi$ | FCC |
| 8.47 ±0.51 | 22.80 ±1.12 | 23.19 ±0.81 | 23.10 ±0.44 | 22.44 ±0.20 | $Al_{0.37}CoCrFeNi$ | FCC |
| 9.36 ±0.12 | 22.25 ±0.36 | 22.51 ±0.78 | 22.41 ±0.78 | 23.47 ±0.62 | $Al_{0.41}CoCrFeNi$ | FCC+BCC |
| 10.82 ±0.04 | 22.55 ±0.16 | 22.54 ±0.06 | 21.55 ±0.57 | 22.55 ±0.61 | $Al_{0.48}CoCrFeNi$ | FCC+BCC |
| 11.55 ±0.27 | 22.33 ±0.53 | 21.79 ±0.38 | 22.03 ±0.99 | 22.29 ±0.28 | $Al_{0.52}CoCrFeNi$ | FCC+BCC/B2 |
| 11.83 ±0.16 | 20.32 ±0.27 | 21.92 ±0.98 | 21.83 ±1.86 | 22.18 ±0.61 | $Al_{0.54}CoCrFeNi$ | FCC+BCC/B2 |
| 14.79 ±0.08 | 20.87 ±0.39 | 22.02 ±0.21 | 21.49 ±0.96 | 20.84 ±0.67 | $Al_{0.69}CoCrFeNi$ | FCC+BCC/B2 |
| 14.79 ±0.78 | 20.32 ±0.27 | 22.70 ±0.33 | 21.65 ±0.67 | 20.55 ±0.32 | $Al_{0.69}CoCrFeNi$ | FCC+BCC/B2 |
| 15.03 ±1.29 | 20.57 ±0.80 | 21.80 ±0.78 | 21.67 ±0.38 | 20.95 ±0.84 | $Al_{0.71}CoCrFeNi$ | FCC+BCC/B2 |
| 16.36 ±0.87 | 21.12 ±0.35 | 21.18 ±0.37 | 21.80 ±0.57 | 19.55 ±0.52 | $Al_{0.78}CoCrFeNi$ | FCC+BCC/B2 |
| 17.09 ±0.36 | 20.25 ±0.80 | 21.48 ±0.78 | 20.59 ±0.36 | 20.61 ±0.57 | $Al_{0.82}CoCrFeNi$ | FCC+BCC/B2 |
| 17.12 ±0.38 | 20.31 ±0.33 | 21.46 ±0.67 | 20.72 ±0.30 | 20.40 ±0.32 | $Al_{0.83}CoCrFeNi$ | FCC+BCC/B2 |
| 17.45 ±1.01 | 20.30 ±0.81 | 21.27 ±0.78 | 20.86 ±0.36 | 20.13 ±0.57 | $Al_{0.85}CoCrFeNi$ | FCC+BCC/B2 |
| 17.63 ±1.24 | 20.22 ±0.90 | 21.21 ±0.79 | 20.73 ±0.39 | 20.22 ±1.19 | $Al_{0.86}CoCrFeNi$ | FCC+BCC/B2 |
| 18.75 ±0.17 | 20.69 ±0.39 | 20.04 ±0.21 | 19.84 ±0.96 | 20.67 ±0.67 | $Al_{0.92}CoCrFeNi$ | FCC+BCC/B2 |
| 20.98 ±0.21 | 19.87 ±1.45 | 19.87 ±1.17 | 19.73 ±1.18 | 19.55 ±1.34 | $Al_{1.06}CoCrFeNi$ | FCC+BCC/B2 |
| 20.97 ±0.12 | 19.33 ±0.06 | 19.46 ±0.99 | 19.81 ±0.38 | 20.42 ±0.29 | $Al_{1.06}CoCrFeNi$ | FCC+BCC/B2 |
| 22.44 ±0.37 | 19.71 ±0.13 | 18.96 ±0.12 | 19.18 ±0.17 | 19.71 ±0.55 | $Al_{1.16}CoCrFeNi$ | BCC/B2 |
| 24.78 ±1.18 | 19.40 ±0.80 | 18.54 ±0.78 | 18.45 ±0.38 | 18.82 ±0.84 | $Al_{1.32}CoCrFeNi$ | BCC/B2 |

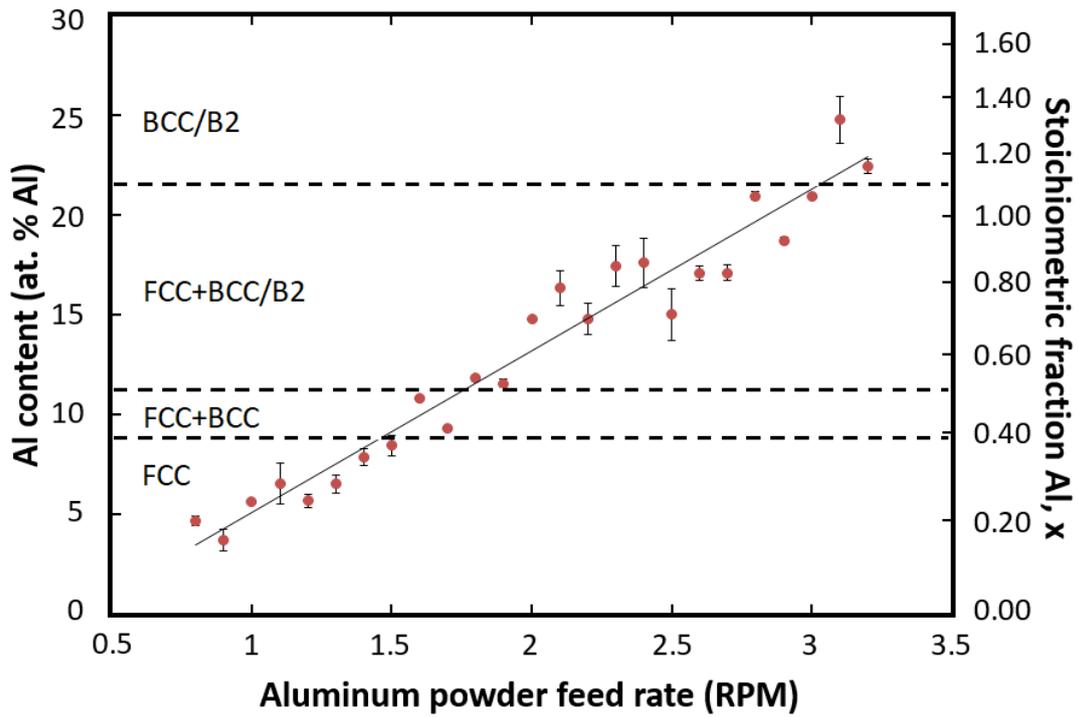

**Figure 1.** The composition range of the laser deposited Al$_x$CoCrFeNi alloy library, as a function of the powder feed rate, determined by EDS. Each point corresponds to the average of four measurements within the deposited patch. Error bars indicate the standard deviation. The observed FCC, BCC/B2, and multi-phase regimes are indicated.

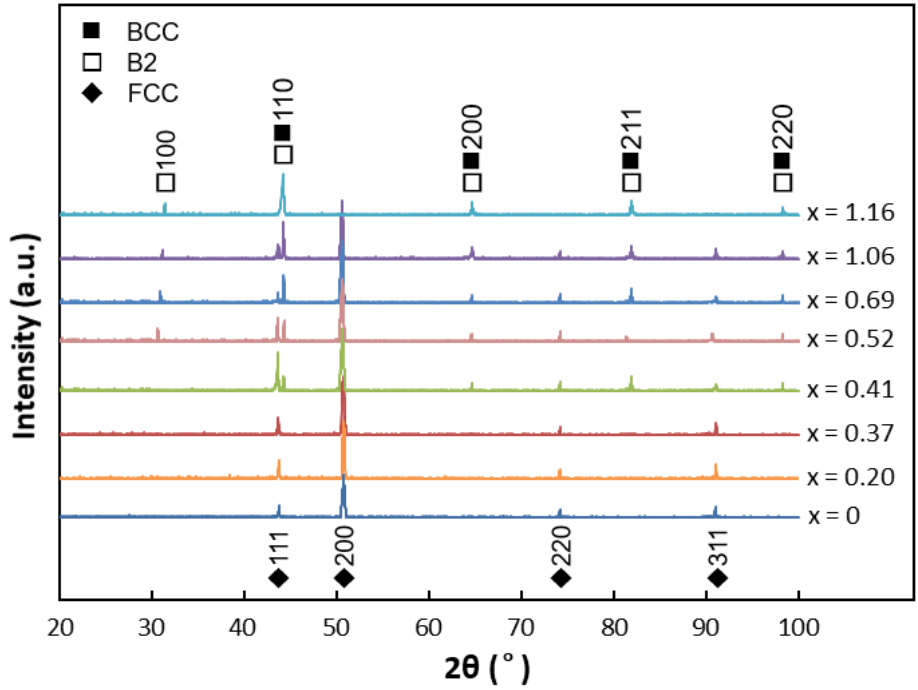

**Figure 2.** X-ray diffraction data for selected Al$_x$CoCrFeNi patches within the alloy library. An FCC structure is observed at low Al contents. With increasing Al content, a disordered BCC structure (x = 0.41) and ordered B2 structure (x = 0.52) form. The FCC structure is no longer observed for x ≥ 1.16.

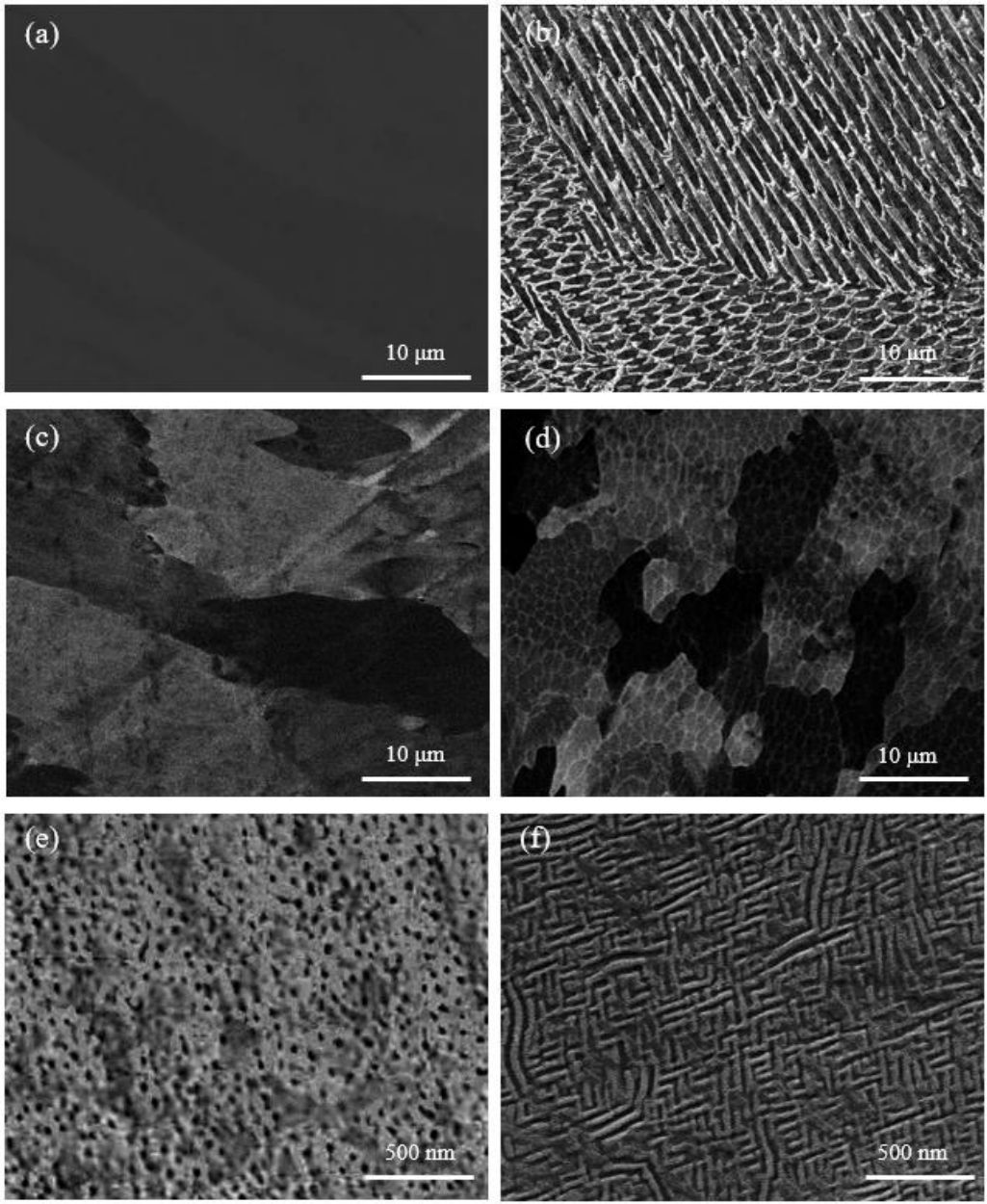

**Figure 3.** Scanning electron micrographs of the microstructural evolution of the laser processed Al$_x$CoCrFeNi alloys with increasing Al content. Secondary electron images of the predominantly FCC structures are shown for (a) x = 0.37 and (b) x = 0.52. Note the appearance of the lighter BCC/B2 phase in (b), forming cell wall-like structures within the main FCC dendrites. Back-scattered electron images of the predominantly BCC/B2 structures are shown for (c) x = 0.83 and (d) x = 1.16. Again, a dendritic/cell wall-type structure is apparent within the grains. Secondary electron images taken (e) within the dendrites and (f) within the interdendritic walls for x = 1.16 show that both regions are dual phase. Spherical nanoparticles are observed within the dendrites, while the interdendritic walls exhibit weave-like structures characteristic of spinodal decomposition.

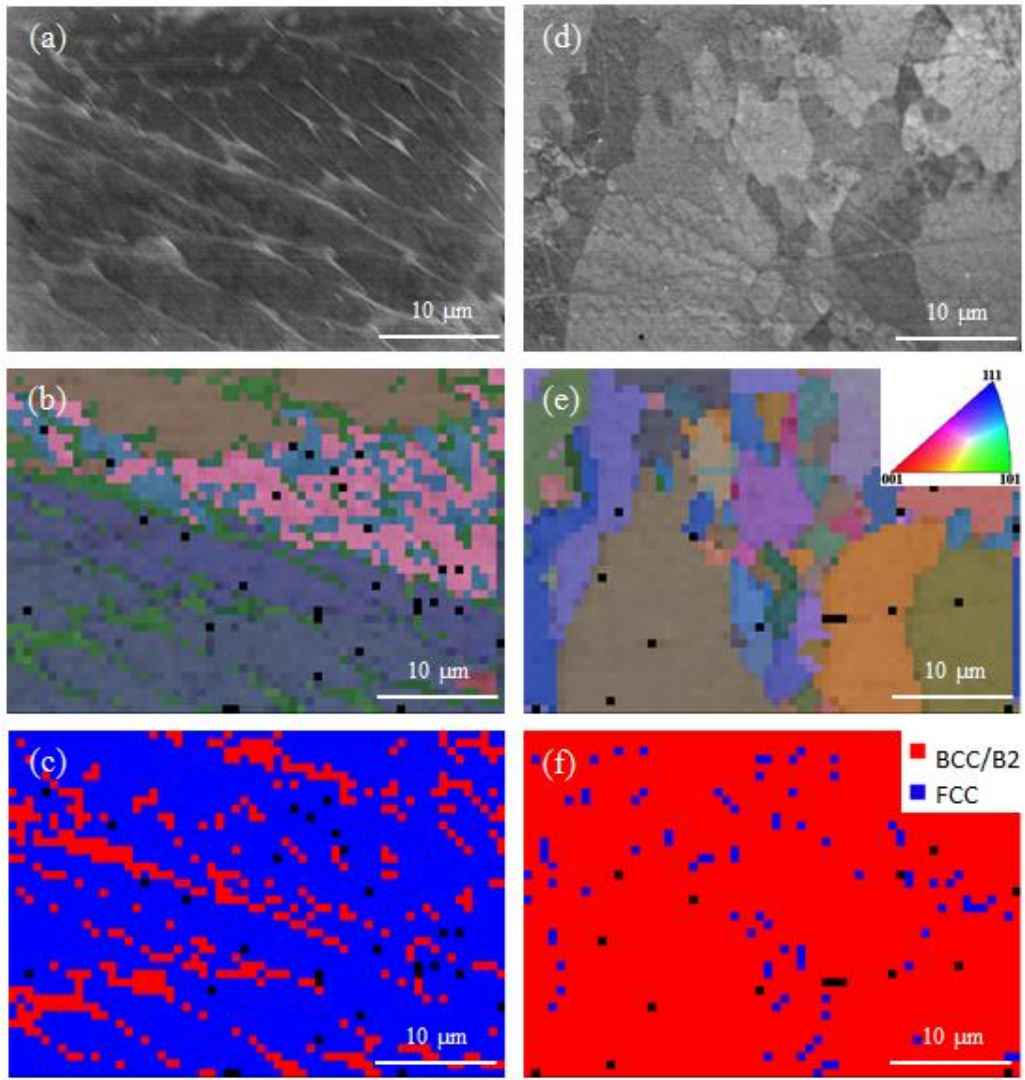

**Figure 4.** Scanning electron micrographs comparing (a-c) primarily FCC Al$_{0.41}$CoCrFeNi and (d-f) primarily BCC/B2 Al$_{0.69}$CoCrFeNi. (a, d) Secondary electron images; (b, e) corresponding EBSD crystal orientation maps; (c, f) corresponding phase maps. Note that the B2 structure could not be differentiated from BCC in the EBSD maps. For low Al contents, the dendrites are FCC and the interdendritic regions are BCC, while at higher Al contents, both the dendrites and interdendritic regions are BCC/B2, and the retained FCC phase is primarily found at the grain boundaries.

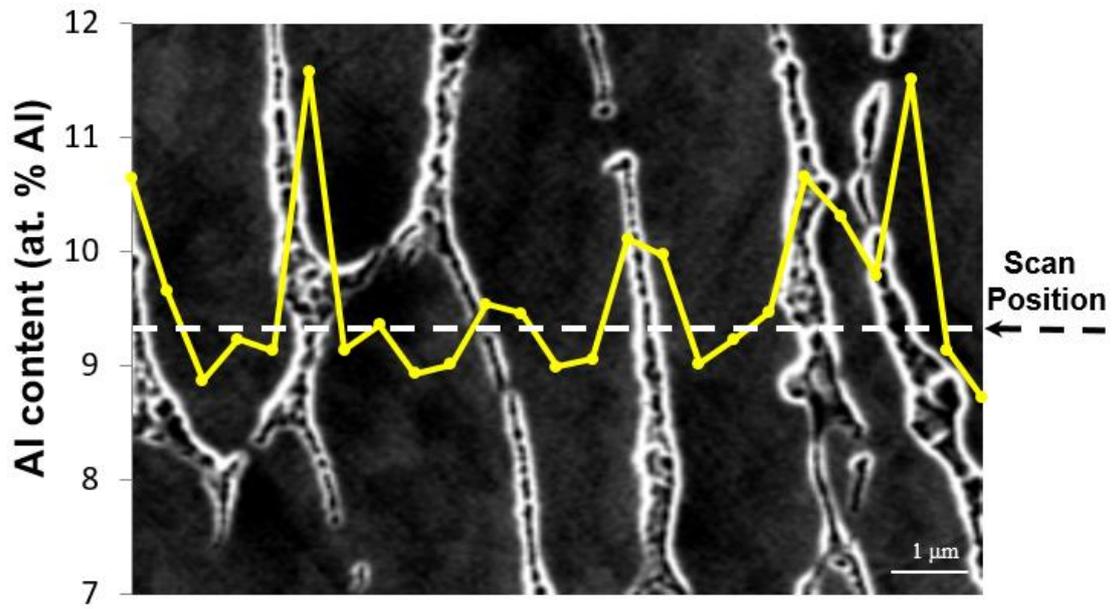

**Figure 5.** Scanning electron micrograph with an overlaid EDS line scan across the FCC/BCC boundaries in Al$_{0.41}$CoCrFeNi, showing compositional segregation into Al-depleted FCC dendrites and Al-enriched BCC regions along the boundaries.

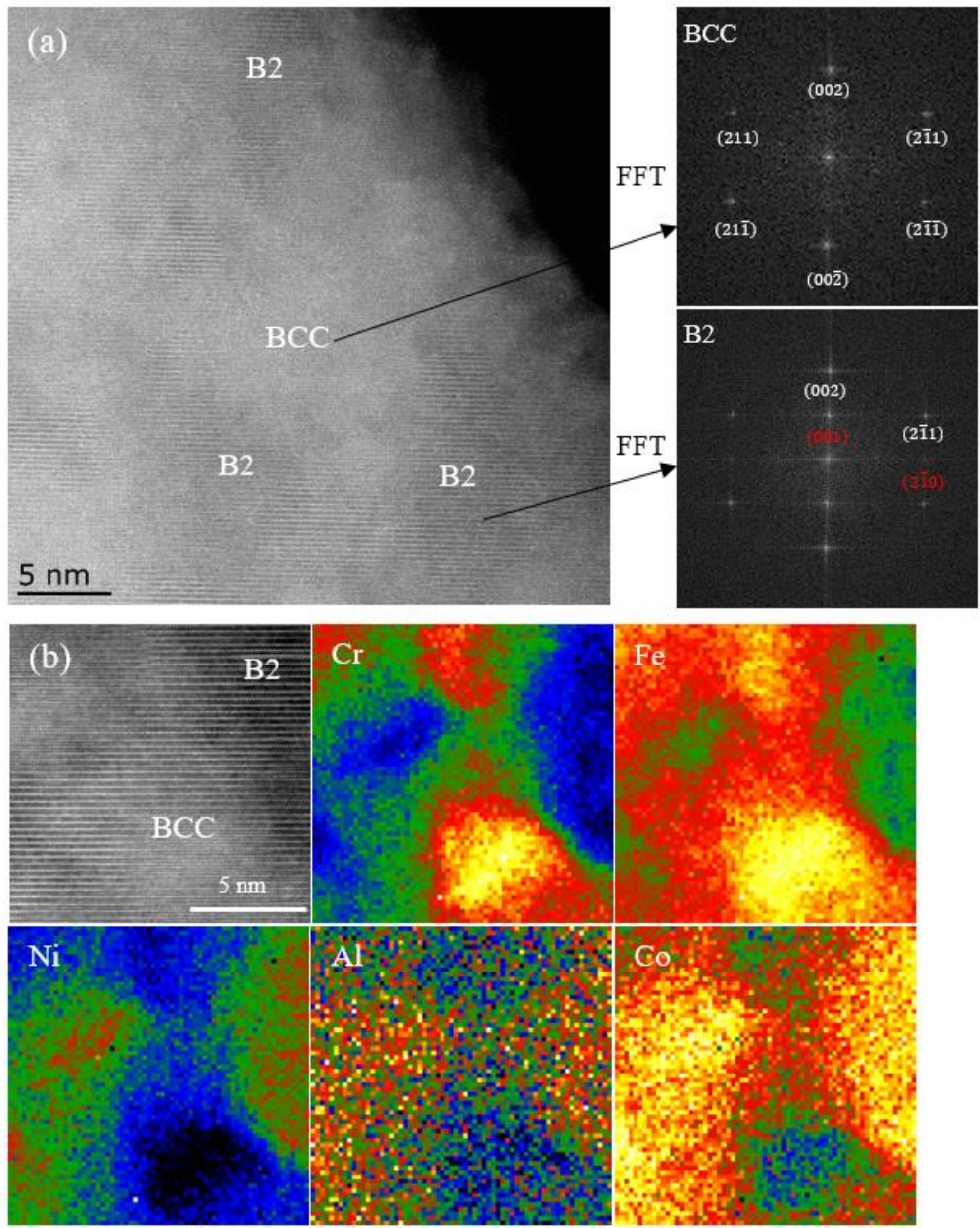

**Figure 6.** (a) HAADF Z-contrast STEM image of Al$_{0.62}$CoCrFeNi, showing phase separation into coherent BCC and B2 regions. The zone axis is [120]. (b) EELS mapping of Al$_{0.62}$CoCrFeNi and corresponding ADF images, showing elemental segregation as heat maps for each element. In comparison to the disordered BCC phase, the ordered B2 structure is rich in Al, Ni and Co but depleted in Cr and Fe.

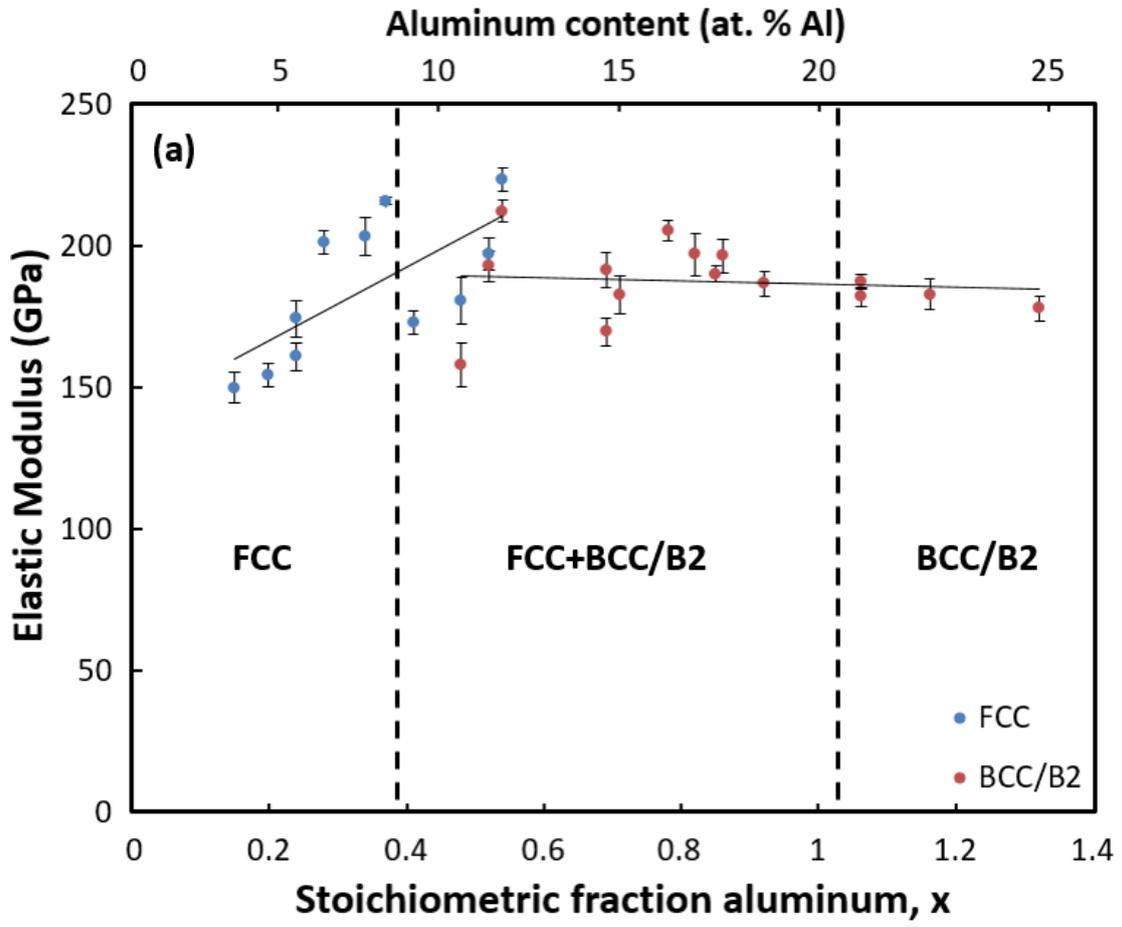

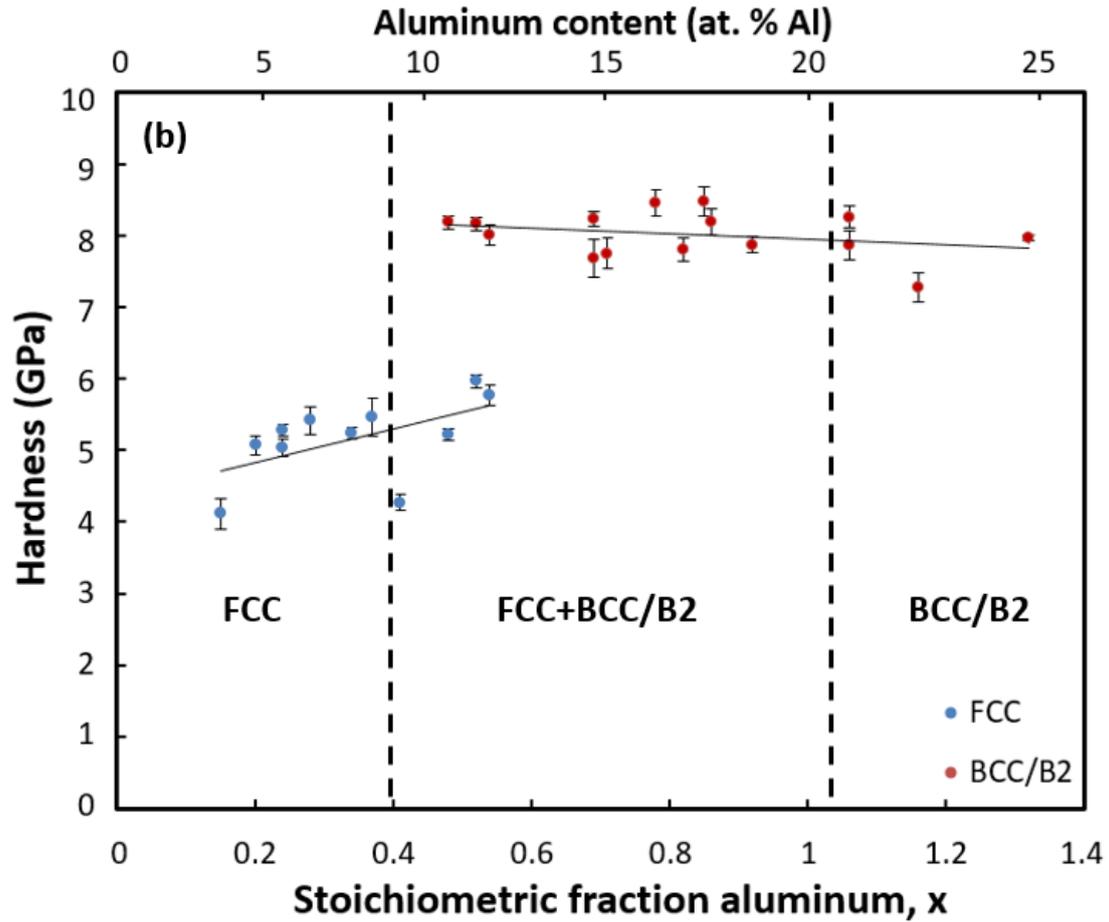

**Figure 7.** (a) Elastic modulus and (b) hardness measured via nanoindentation of the FCC and BCC/B2 structures in the Al$_x$CoCrFeNi compositional library, as a function of the nominal stoichiometric fraction of Al in each library patch. Each point represents the average of four measurements. Error bars represent the standard deviation. Dashed lines indicate the composition ranges where the microstructure is predominantly FCC (low x), predominantly BCC/B2 (high x), and multiphase.

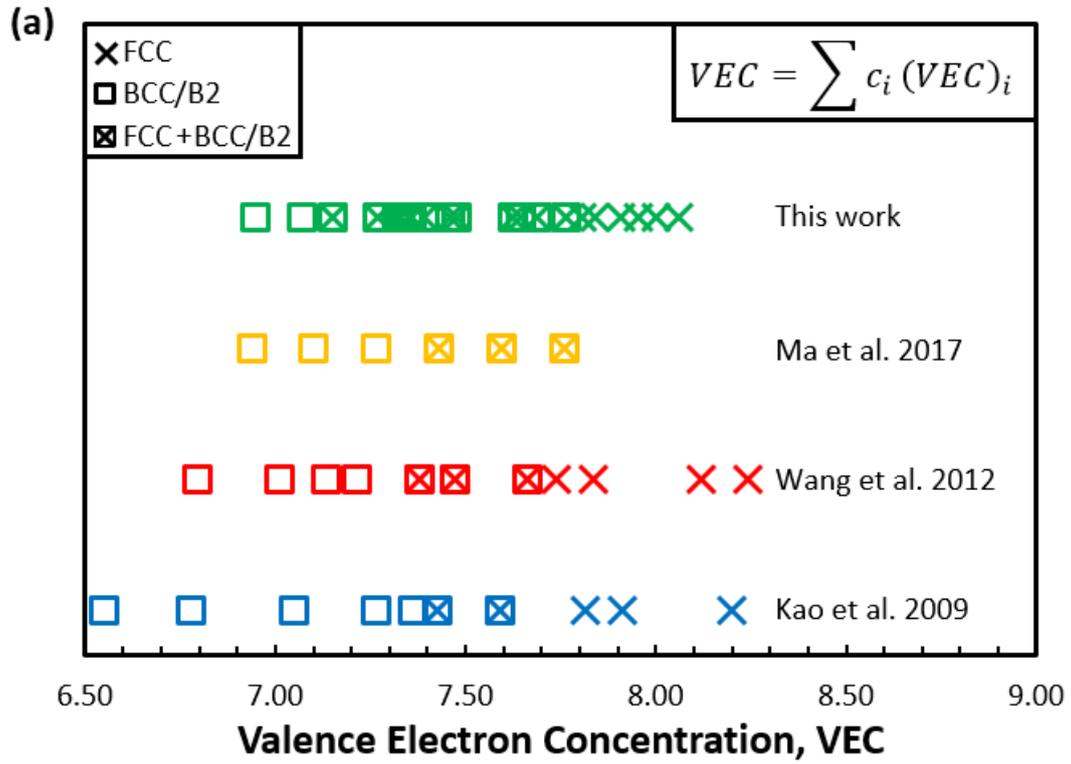

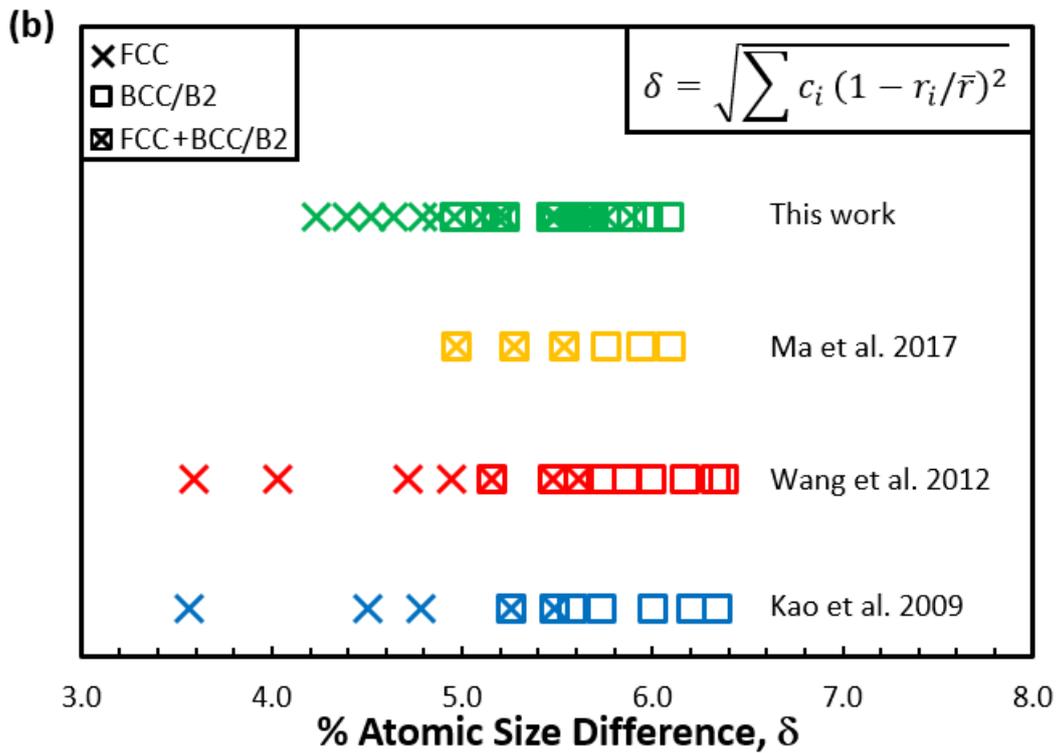

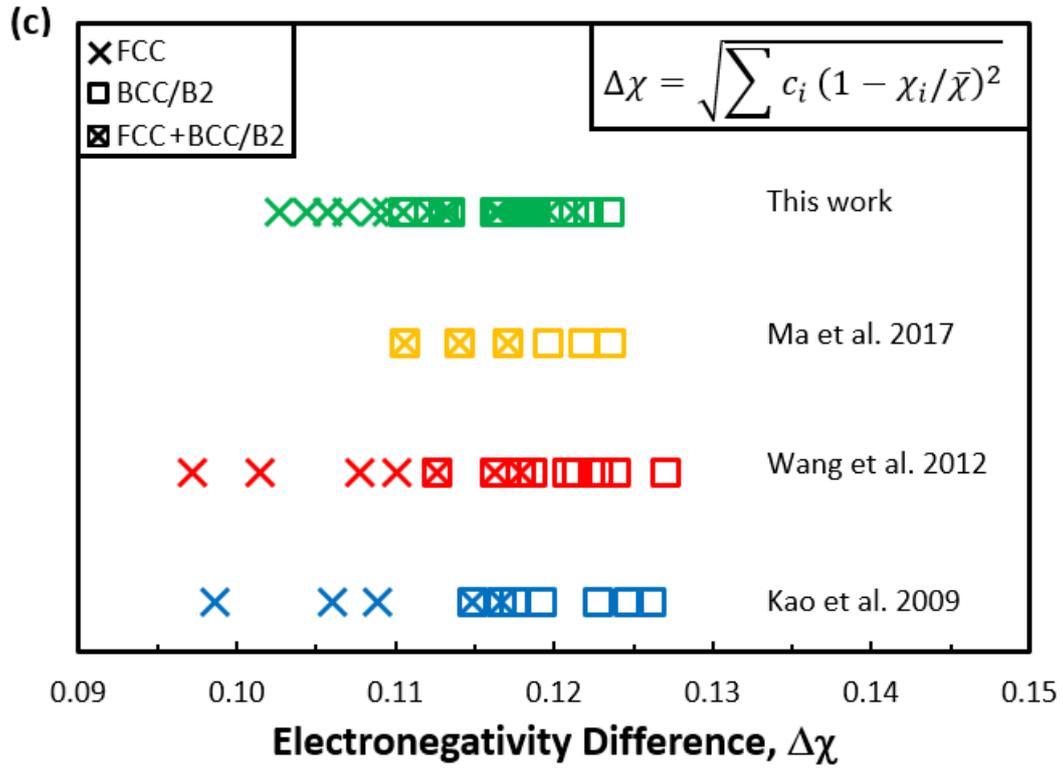

**Figure 8.** Correlation of FCC and BCC/B2 phase formation with (a) valence electron concentration, (b) % difference in atomic size, and (c) electronegativity difference, as defined in [20], for the laser processed $Al_xCoCrFeNi$ alloys in the present compositional library, compared with the cast alloys investigated in other work [11,13,14].